\documentclass[a4paper,prd,superscriptaddress,twocolumn,aps,notitlepage,floatfix]{revtex4}
\usepackage{amsmath, amssymb, amsfonts, mathtext}
\usepackage[usenames]{color}
\usepackage{graphicx}
\usepackage[pdftex,unicode,colorlinks,citecolor=blue,linkcolor=blue,
bookmarks=true,bookmarksopen=true,bookmarksopenlevel=3,bookmarksnumbered=true]{hyperref}

\renewcommand{\Re}{\mathop{\rm Re}\nolimits}
\newcommand{\hence}{\ \Rightarrow\ }

\begin{document}

\title{Paired carriers as a way to reduce quantum noise of multi-carrier gravitational-wave detectors}

\author{Mikhail Korobko}
\affiliation{Institut f{\"u}r Gravitationsphysik, Leibniz Universit\"at Hannover and Max-Planck-Institut f\"ur Gravitationsphysik
(Albert-Einstein-Institut), Callinstr. 38, 30167 Hannover, Germany}
\author{Nikita Voronchev}
\affiliation{Faculty of Physics, Moscow State University, Moscow 119991, Russia}
\author{Haixing Miao}
\affiliation{School of Physics and Astronomy,
University of Birmingham, Birmingham, B15 2TT, United Kingdom}
\author{Farid Ya. Khalili}
\affiliation{Faculty of Physics, Moscow State University, Moscow 119991, Russia}

\begin{abstract}
  We explore new regimes of laser interferometric gravitational-wave detectors with multiple optical carriers which allow to reduce the quantum noise of these detectors. In particular, we show that using two carriers with the opposite detunings, homodyne angles, and squeezing angles, but identical other parameters (the antisymmetric carriers), one can suppress the quantum noise in such a way that its spectrum follows the Standard Quantum Limit (SQL) at low frequencies. Relaxing this antisymmetry condition, it is also possible to slightly overcome the SQL in broadband. Combining several such pairs in the xylophone configuration, it is possible to shape the quantum noise spectrum flexibly.
\end{abstract}

\maketitle


\section{Introduction}

Currently, the second generation large-scale laser interferometric gravitational-wave (GW) detectors: Advanced LIGO \cite{AdvLIGOsite, Harry2010}, Advanced VIRGO \cite{AdvVIRGOsite, Acernese2006-2}, and KAGRA \cite{KAGRAsite, Kanda2011} are under construction. In particular, construction of two Advanced LIGO interferometers is almost complete and they will start to gather scientific data soon. Sensitivities of these detectors are expected to be limited by the quantum noise. Namely, at higher frequencies the shot noise will dominate, originating from quantum fluctuation of the phase of the optical field inside the interferometer. At lower frequencies, the radiation pressure noise created by the amplitude fluctuations will constitute the major part of the noise budget. The shot noise is inversely proportional to the optical power circulating inside the interferometer, while the radiation pressure noise is proportional to it\,\cite{Caves1981} --- the optimal point where these two noises are equal to each other is known as the Standard Quantum Limit (SQL)\,\cite{92BookBrKh}.

It has to be emphasized that the SQL represents an ultimate sensitivity limit only for a simplest class of position measurement schemes, which, however, encompasses the baseline design of all second generation GW detectors (see details below in Sec.\,\ref{sec:S_sum}). Several methods of overcoming this limit suitable for the laser GW detectors were proposed (see {\it e.g.}\,the review paper \cite{12a1DaKh}; we discuss briefly two most well known ones in Sec.\,\ref{sec:S_sum}). In most cases, they require significant modifications in the interferometer design; and in order to take full advantage of these methods, the other noise sources of non-quantum origin (so-called technical noise) have to be suppressed correspondingly. Due to the these reasons, these configurations typically are considered as possible candidates for implementation only in the planned third-generation GW detectors \cite{Miao1305_3957}, like the Einstein Telescope \cite{ETsite, Hild_CQG_28_094013_2011, Sathyaprakash2012} or the LIGO III \cite{wp2014}, where the technical noise will be reduced by about one order of magnitude (in comparison to the second generation detectors). In particular, a so-called {\it xylophone} configuration is planned for the Einstein Telescope \cite{Hild2010, Hild_CQG_28_094013_2011}, which consists of two independent interferometers, optimized for low-frequency and high-frequency GW signals, respectively.

However, in the planned Advanced LIGO noise budget, there is a quite large margin between the quantum noise and the technical noise in the low-frequency band $10\,\text{-}\,50\, {\rm Hz}$ \cite{Harry2010}, opening the opportunity to improve the sensitivity in this important frequency band by using one a simplified form of one of the above mentioned methods. In particular, the injection of frequency-dependent squeezed light created by means of a single relatively short (16m) filter cavity (a simplified form of the {\it pre-filtering} topology proposed in \cite{02a1KiLeMaThVy}) is considered as a very probable option for upgrading during some later stage of the Advanced LIGO \cite{Evans_PRD_88_022002_2013}.

Another approach to reducing quantum noise in GW detectors is modification of the test masses' dynamics by means of the {\it optical spring} effect which arises in the detuned interferometers \cite{99a1BrKh, 01a2Kh, Buonanno2002}. The optical springs convert GW detectors test masses into harmonic oscillators with eigenfrequencies within the detection band (rigorously speaking, this approach does not allow to overcome the SQL, but instead reduces the SQL itself around the eigenfrequency). Unfortunately, the optical springs allow to improve the sensitivity in a limited frequency band, while substantially degrading it at other frequencies.

A further development of this method was proposed in papers \cite{11a1KhDaMuMiChZh, 12a1eDaKhVo}. It is based on use of two optical carriers which create two optical springs of the opposite signs. Provided the appropriate power, detuning and bandwidth of the carriers, the total effect of the double optical spring can be described as a {\it negative optical inertia}. It cancels the positive inertia of the test masses, thus increasing their response to gravitational waves and correspondingly reducing the SQL within a broad band from zero frequency to some upper frequency limited by the available optical power. Unfortunately, estimates show that for parameters planned for the Advanced LIGO, this upper frequency is equal to only $\sim50\,{\rm Hz}$, and scales very slowly (as $I_c^{1/3}$) with the circulating optical power $I_c$ \cite{12a1eDaKhVo}.

\begin{figure}
  \includegraphics[width=\columnwidth]{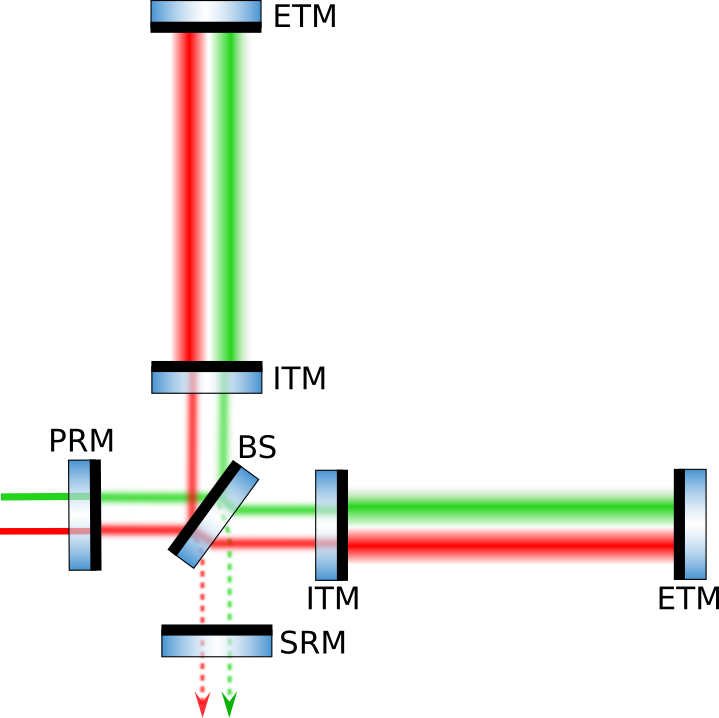}
  \caption{Scheme of a second generation laser GW detector with two carriers.}\label{fig:scheme}
\end{figure}

In the articles \cite{Corbitt2007, Rehbein2008} the double-carrier configuration was proposed as a mean to create a dynamically stable optical spring \footnote{It is known, that depending on the detuning sign, a single carrier creates either positive rigidity accompanied by negative damping, or negative rigidity with positive damping. Both cases are evidently unstable. However, combining two carriers with different powers and detunings, it is possible to implement the stable configuration with the positive total rigidity and positive total damping.}. The scheme considered in \cite{Rehbein2008} is shown in Fig.\,\ref{fig:scheme}. In essence, this is the standard Michelson/Fabry-P\'erot topology of the second generation GW detectors, but with two optical pump sources, which either have to have orthogonal polarizations, or have to be separated by one or more FSRs of the interferometer, in order to avoid interference between them. Each of the two output beams is supposed to be measured by its own homodyne detector, and their output signals are combined with the optimal weight functions.

In addition, the so called {\it annihilation} regime was considered in \cite{Rehbein2008},  which uses the two carriers with equal power and opposite detunings; as a result, the optical springs created by these two carriers completely cancel each other. Here, we analyze this regime in more detail and show, that it allows to reduce the radiation pressure noise in the second generation GW detectors down to the level of their technical noise. We show also, that using a several such pairs, it is possible to implement the xylophone configuration within the single interferometer.

We assume in this paper, that the main parameters of the interferometer correspond to the ones planned for the Advanced LIGO \cite{Harry2010}, see Table\,\ref{tab:notations}. In particular, we suppose, that the total circulating optical power of the all carriers is limited to $840\,{\rm kW}$, which corresponds to the normalized power $J=(2\pi\times100)^3\,{\rm s}^{-3}$ (the main notations used throughout this paper are listed in Table\,\ref{tab:notations}). We suppose also that for each carrier, a frequency-independent squeezed light can be injected into the dark port of the interferometer as it was proposed by C.~Caves in \cite{Caves1981}.


This paper is organized as follows. In the next section we briefly review the main features of quantum noise in the GW detectors. In Sec.\,\ref{sec:idea} we analyze the main features of the multi-carrier quantum noise. In Sec.\,\ref{sec:optimization}, we present the results of the numerical optimization of this noise. In Sec.\,\ref{sec:discussion}, we discuss the main advantages and disadvantages of the proposed method and the prospects of its use in future GW detectors. In the Appendix, the effective quantum noise spectral densities for the multi-carrier configuration are calculated. 
\begin{table}
  \begin{ruledtabular}
    \begin{tabular}{ll}
      Quantity & Description \\ \hline
      $c$               & Speed of light \\
      $\hbar$           & Reduced Plank constants \\
      $M = 40\,{\rm kg}$  & Mass of each of the arm cavities mirrors \\
      $L = 4\,{\rm km}$ & Length of the interferometer arm cavities \\
      $\omega_p = 2\pi c/1.064\,\mu{\rm m}$ & Optical pump frequency \\
      $\omega_o$        & Resonance frequency of the interferometer \\
      $\gamma$          & Half-bandwidth of the interferometer \\
      $\delta = \omega_p-\omega_o$ & Detuning \\
      $\varGamma=\sqrt{\gamma^2+\delta^2}$ & Effective half-bandwidth \\
      $\beta = \arctan\dfrac{\delta}{\gamma}$ & Normalized detuning \\
      $\Omega$          & Audio sideband frequency of the GW signal \\
      $I_c$             & Optical power circulating in the arm cavities \\
      $J = \dfrac{4\omega_pI_c}{MLc}$ & Normalized optical power \\
      $\zeta$           & Homodyne angle \\
      $e^{2r}$          & Squeezing power \\
      $\theta$          & Squeezing angle \\
      $\eta$            & Unified quantum efficiency
    \end{tabular}
  \end{ruledtabular}
  \caption{Main notations used in this paper.}\label{tab:notations}
\end{table}

\section{General structure of quantum noise}\label{sec:S_sum}

In the particular case of the unmodified free mass mechanical dynamics (without the optical springs), which we consider in this paper, spectral density of quantum noise of the laser interferometric GW detectors, normalized to GW strain, is equal to (see {\it e.g.}\,\cite{12a1DaKh}):
\begin{equation}\label{eq:s_sum_1}
  S_{\rm sum}(\Omega) = \frac{8}{L^2}\biggl[
      S_{xx}(\Omega) - \frac{2\Re S_{xF}(\Omega)}{M\Omega^2}
      + \frac{S_{FF}(\Omega)}{M^2\Omega^4}
    \biggr] ,
\end{equation}
where $S_{xx}(\Omega)$, $S_{FF}(\Omega)$, and $S_{xF}(\Omega)$ are, respectively, spectral densities of the shot noise, the radiation pressure noise, and the cross-correlation spectral density of these two noises, which obey the following uncertainty relation:
\begin{equation}\label{eq:uncert}
  S_{xx}(\Omega)S_{FF}(\Omega) - |S_{xF}(\Omega)|^2 = \frac{\hbar^2}{4\eta(\Omega)} \,,
\end{equation}
where $\eta \le 1$ is the quantum efficiency of the detector, which takes into account both the optical losses and the photodetector non-unity quantum efficiency. For simplicity, we will assume the ideal case of $\eta=1$ in the rest of this section (as we show later, the optical losses significantly influence the sensitivity of the method which we consider in this paper; however, they are not important for understanding of the basic features of the quantum noise).

Suppose first that the shot noise and the radiation pressure noise are uncorrelated: $S_{xF}(\Omega)=0$. In this case the minimum of \eqref{eq:s_sum_1} is achieved by
\begin{equation}\label{S_FF_nocorr}
  S_{FF}(\Omega) = \frac{\hbar M\Omega^2}{2}
\end{equation}
and is equal to the free mass SQL:
\begin{equation}\label{S_SQL}
  S_{\rm SQL}(\Omega) = \frac{8\hbar}{L^2M\Omega^2}
\end{equation}

In the general case of $S_{xF}\ne0$, the minimum of \eqref{eq:s_sum_1} [with account of  the condition \eqref{eq:uncert}] is given by
\begin{equation}\label{S_xF_corr}
  S_{xF}(\Omega) = \frac{S_{FF}(\Omega)}{M\Omega^2} \,,
\end{equation}
and is equal to
\begin{equation}\label{S_min_corr}
  S_{\rm opt}(\Omega) = \frac{2\hbar^2}{L^2S_{FF}(\Omega)} \,.
\end{equation}
Therefore, using the cross-correlation of the shot noise and the radiation pressure noise, it is possible to achieve arbitrary high sensitivity, providing $S_{FF}$ is sufficiently large, that is, the optical power is sufficiently strong.

In the laser interferometric GW detectors, the cross correlation can be introduced relatively easy by means of a homodyne detection with an optimized homodyne angle $\zeta$. However, in order to reach or overcome the SQL in a finite frequency band, the quantum noise components have to have within this band the proper frequency dependencies dictated by Eqs.\,\eqref{S_FF_nocorr} or \eqref{S_xF_corr}, respectively.

Consider the important example of the resonance-tuned interferometer ($\delta=0$); it is this case is planned for the second generation GW detectors. In order to avoid unnecessary complication, we also suppose here that squeezed light is not used (however, the squeezing will be taken into account below in Sections \ref{sec:idea} and \ref{sec:optimization}).

If its shot and radiation pressure noises are uncorrelated then the corresponding total quantum noise spectral density is equal to (see \cite{12a1DaKh})
\begin{equation}\label{S_sum_base}
  S_{\rm sum}(\Omega) = \frac{S_{\rm SQL}(\Omega)}{2}
    \biggl[\frac{1}{\mathcal{K}_{\rm PM}(\Omega)} + \mathcal{K}_{\rm PM}(\Omega)\biggr] ,
\end{equation}
where
\begin{equation}
  \mathcal{K}_{\rm PM}(\Omega) = \frac{2J\gamma}{\Omega^2(\gamma^2+\Omega^2)}
\end{equation}
is the optomechanical coupling factor of the position meter \cite{02a1KiLeMaThVy}. It is easy to see that the spectral density \eqref{S_sum_base} reaches the SQL only at one frequency which satisfies the following equation:
\begin{equation}
  \Omega^2(\gamma^2+\Omega^2) = 2J\gamma \,,
\end{equation}
and goes above the SQL at all other frequencies. In the rest of this paper, this particular case will be referred to as the {\it baseline interferometer}. We will draw this spectral density as the reference in all plots below, for the particular case of $J=(2\pi\times100)^3\,{\rm s}^{-3}$ and $\gamma=2\pi\times500\,{\rm s}^{-1}$, which approximately corresponds to the values planned for the Advanced LIGO \cite{Harry2010}.

Then consider the case of $S_{xF}\ne0$. The structure of equation \eqref{S_xF_corr} suggests that this equation can be fulfilled in a broad band by making either $S_{FF}$ or $S_{xF}$ frequency dependent. These two options correspond to two methods of overcoming the SQL considered as the most probable candidates for implementation in the third generation GW detectors. The first one, proposed in the work \cite{02a1KiLeMaThVy}, is based on use of additional {\it filter cavities}, which allow to create the frequency-dependent cross-correlation of the quantum noises.

The second method which is more relevant for our consideration, so-called ``quantum speedmeter'', was first proposed as semi-gedanken scheme in \cite{90a1BrKh} and later developed into two realistic interferometer topologies (based on the Sagnac interferometer and on the ordinary Michelson one, but with an additional {\it sloshing cavity}) in papers \cite{00a1BrGoKhTh, Purdue2001, Purdue2002, Chen2002, 04a1Da}. This scheme is sensitive to the velocity of test masses, instead of their displacement (hence the designation ``speedmeter''). This corresponds to the following characteristic frequency dependencies of the quantum noise spectral densities:
\begin{align}
  S_{xx}(\Omega) &= \frac{S_{vv}}{\Omega^2} \,, &
  S_{FF}(\Omega) &= \Omega^2S_{pp} \,,
\end{align}
where $S_{vv}$, $S_{pp}$ are spectral densities of the velocity measurement noise and the momentum perturbation noise, respectively \footnote{Note that in the quantum speedmeter scheme, the effective coupling of the test mass with the meter is proportional to the velocity $v$ of the former one; therefore its momentum $p \ne mv$ and $S_{pp}\ne m^2S_{vv}$}. Within the bandwidth of the interferometer, $\Omega<\gamma$, these spectral densities can be considered as frequency independent ones, which allows to fulfill conditions (\ref{S_FF_nocorr}, \ref{S_xF_corr}) in broadband by measuring a proper homodyne angle and without filter cavities.

The explicit equation for the total quantum noise spectral density of the speedmeter is the following \cite{12a1DaKh}:
\begin{multline}\label{S_sum_SM}
  S_{\rm sum}(\Omega) = \frac{S_{\rm SQL}(\Omega)}{2}\biggl[
      \frac{1}{\mathcal{K}_{\rm SM}(\Omega)\sin^2\zeta} - 2\cot\zeta \\
      + \mathcal{K}_{\rm SM}(\Omega)
    \biggr] ,
\end{multline}
where the optomechanical coupling factor of the speedmeter $\mathcal{K}_{\rm SM}$ is equal to
\begin{subequations}
  \begin{equation}
    \mathcal{K}_{\rm SM}(\Omega) = \frac{4J\gamma}{(\gamma^2+\Omega^2)^2} \,
  \end{equation}
  for the Sagnac-type speedmeter and
  \begin{equation}
    \mathcal{K}_{\rm SM}(\Omega) = \frac{4J\gamma}{4\gamma^4 + \Omega^4} \,
  \end{equation}
\end{subequations}
for the speedmeter realized by using an additional sloshing cavity (only the low-frequency optimized case is shown for brevity, and refer to Ref.\,\cite{Purdue2002} for more details). Note that in both cases (in contrast with $\mathcal{K}_{\rm PM}$), this factor does not depend on $\Omega$ in the asymptotic case of $\Omega\ll\gamma$.

\begin{figure}
  \includegraphics{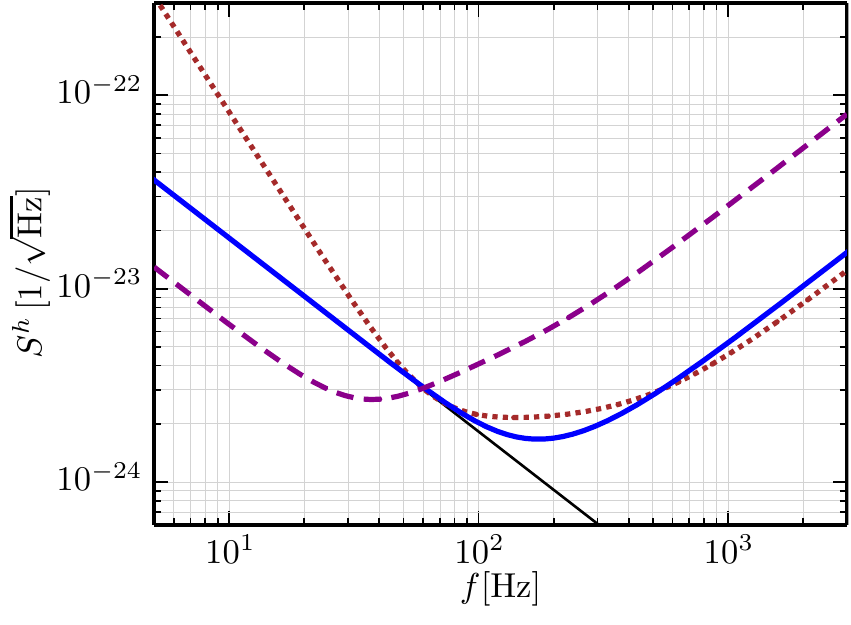}
  \caption{Plots of the total noise spectral densities of: the baseline interferometer \eqref{S_sum_base}, at $\gamma=2\pi\times500\,{\rm s^{-1}}$ (dots); the Sagnac speedmeter without the quantum noises cross-correlation \eqref{S_sum_SM_nocorr}, at $\gamma=2^{2/3}\times2\pi\times100\,{\rm s^{-1}}$ (solid); the Sagnac speedmeter with the cross-correlation \eqref{S_sum_SM_corr}, at $\gamma=2\pi\times100\,{\rm s^{-1}}$, $\cot\zeta=4$ (dashes).  Thin solid line: the SQL \eqref{S_SQL}. In all cases, $J=(2\pi\times100)^3\,{\rm s^{-3}}$ and $\eta=1$ (no losses).}\label{fig:baseline}
\end{figure}

Therefore, if the shot noise and the radiation pressure noise are not correlated, that is $\zeta=\pi/2$, then the low-frequency optimization
\begin{equation}
  \mathcal{K}_{\rm SM}(0) = 1 \hence J = \frac{\gamma^3}{4}
\end{equation}
gives the total noise spectral density that asymptotically follows the SQL within the interferometer bandwidth. In particular, in the Sagnac speedmeter case, it is equal to
\begin{multline}\label{S_sum_SM_nocorr}
  S_{\rm sum}(\Omega) = \frac{S_{\rm SQL}(\Omega)}{2}\biggl[
      \frac{\gamma^4}{(\gamma^2+\Omega^2)^2} + \frac{(\gamma^2+\Omega^2)^2}{\gamma^4}
    \biggr] .
\end{multline}
In contrast, using the quantum noise cross correlation at low frequencies by choosing:
\begin{equation}
  \cot\zeta = \mathcal{K}_{\rm SM}(0) = \frac{4J}{\gamma^3}
\end{equation}
gives the total noise spectral density below the SQL within the interferometer bandwidth:
\begin{multline}\label{S_sum_SM_corr}
  S_{\rm sum}(\Omega) = \frac{S_{\rm SQL}(\Omega)}{2\mathcal{K}_{\rm SM}(\Omega)}\Bigl\{
      1 + \bigl[\mathcal{K}_{\rm SM}(0) -  \mathcal{K}_{\rm SM}(\Omega)\bigr]^2
    \Bigr\} .
\end{multline}
These two scenarios are illustrated in Fig.\,\ref{fig:baseline}, where the spectral densities (\ref{S_sum_base}, \ref{S_sum_SM_nocorr}, \ref{S_sum_SM_corr}) are plotted for some characteristic values of $\gamma$ and $J$.

Now, having discussed briefly the quantum noise of the single-carrier interferometers, we are in position to introduce the quantum noise for multiple carriers.

\section{Multi-carrier shaping of quantum noise}\label{sec:idea}

\subsection{Speedmeter-like shot noise in Michelson/Fabry-Perot interferometer}

In a general case of an arbitrary detuning $\delta$ and homodyne angle $\zeta$, the quantum noise spectral densities of the ordinary Michelson/Fabry-P\'erot interferometer have sophisticated frequency dependencies, see Eqs.\,\eqref{SxSFSxF}. In particular if
\begin{equation}\label{low_freq}
  \left|\frac{\sin(\zeta-\beta)}{\sin\zeta}\right|\,\varGamma \ll \Omega \ll \varGamma \,,
\end{equation}
then the shot noise spectral density has a speedmeter-type frequency dependence:
\begin{equation}\label{eq:Sxx_speed}
  S_{xx}(\Omega) \propto \frac{1}{\Omega^2} \,.
\end{equation}
However, frequency dependencies of the other two spectral densities are improper: $S_{FF}(\Omega)\propto\Omega^0$ instead of $S_{FF}(\Omega)\propto\Omega^2$ and $S_{xF}(\Omega)\propto1/\Omega$ instead of $S_{xF}(\Omega)\propto\Omega^0$. Moreover, while the quantum speedmeter requires the free mass dynamics, in the detuned interferometer the dynamics of the test masses is modified by the optical rigidity \cite{Buonanno2003}. Therefore, the frequency dependence \eqref{eq:Sxx_speed} by itself does not allow to realize the  speedmeter type total quantum noise.

However, both the cross-correlation and the optical spring can be canceled using the annihilation regime discussed in \cite{Rehbein2008}. Note that: $S_{xx}$ is an even function of $\delta$, $\zeta$, $\theta$; $S_{xF}$ is an odd function of these three parameters; and $K$ is an odd function of $\delta$ [see Eqs.\,(\ref{SxSFSxF}, \ref{K})]. Therefore two carriers with the following parameters:
\begin{subequations}\label{antisymm}
  \begin{align}
    J_1 &= J_2 \,, \\
    r_1 &= r_2 \,, \\
    \varGamma_1 &= \varGamma_2 \,, \\
    \beta_1 &= -\beta_2 \,, \\
    \zeta_1 &= -\zeta_2 \,, \label{zeta_antisymm} \\
    \theta_1 &= -\theta_2 \label{theta_antisymm}
  \end{align}
\end{subequations}
(the {\it antisymmetric} carriers) create the effective position meter with canceled optical spring and with the quantum noise spectral densities equal to [see Eqs.\,\eqref{S_eff} in the Appendix]
\begin{subequations}\label{S_2_carriers}
  \begin{gather}
    S_{xx}^{\rm eff}(\Omega) = \frac{S_{xx}(\Omega)}{2} \,, \label{S_xx_eff2} \\
    S_{FF}^{\rm eff}(\Omega) = \eta(\Omega)\frac{\hbar^2}{4S_{xx}^{\rm eff}(\Omega)}
      + 2[1-\eta(\Omega)]S_{FF}(\Omega)\,, \label{S_FF_eff2} \\
    S_{xF}^{\rm eff}(\Omega) = 0 \label{S_xFF_eff2} \,.
  \end{gather}
\end{subequations}
where $S_{xx}$, $S_{FF}$ describe the individual carriers.

The first (major) term of the back action noise spectral density \eqref{S_FF_eff2}, being proportional to $\Omega^2$, has the proper speedmeter-like frequency dependence. The second one (originating from the optical losses) has the ordinary position meter spectral dependence \eqref{S_F_PM}, which degrades the effect of the described regime.

It is worth noting that the effective back action noise is smaller, than that just the sum of back action noises of the individual carriers, $S_{FF}^{\rm eff}<2S_{FF}$. This means that the effective back action noise actually is a {\it conditional} one, that is, it describes only the residual noise remaining after subtraction of the part known to the observer due to the cross-correlation of the shot noise and the radiation pressure noise. Note that while the residual cross-correlation \eqref{S_FF_eff2} is canceled the weight functions for the individual output signals depend on the cross-correlation spectral densities of the individual carriers, see Eq.\,\eqref{alpha_j_opt}.

Due to the absence of the residual cross-correlation, opposite to the ``real'' speedmeter case of Eq.\,\eqref{S_sum_SM_corr}, the perfectly antisymmetric carriers allow only to reach the SQL in a broad band, but can not overcome it, like in \cite{Purdue2001, Purdue2002, Chen2002, 04a1Da}. However, due to quite moderate margin between the SQL and the low-frequency technical noise  planned for the second generation GW detectors (most notably, the mirrors coating and the suspension thermal noise, and the gravity gradient noise), only very limited low-frequency sensitivity gain can be provided by the ``real'' speedmeter [Eq.\,\eqref{S_sum_SM_nocorr}], while the use of $\zeta\ne\pi/2$ noticeably increases the shot (high-frequency) noise, see Fig.\,\ref{fig:baseline}.

Relaxing in some extent the anti-symmetry condition \eqref{antisymm} by removing constrains for the homodyne and squeezing angles $\zeta$ and $\theta$, it is possible to create the residual cross-correlation $S_{xF}$ and overcome the SQL in some frequency band. We consider this possibility in more detail in Sec.\,\ref{sec:optimization}.

\begin{figure}
  \includegraphics{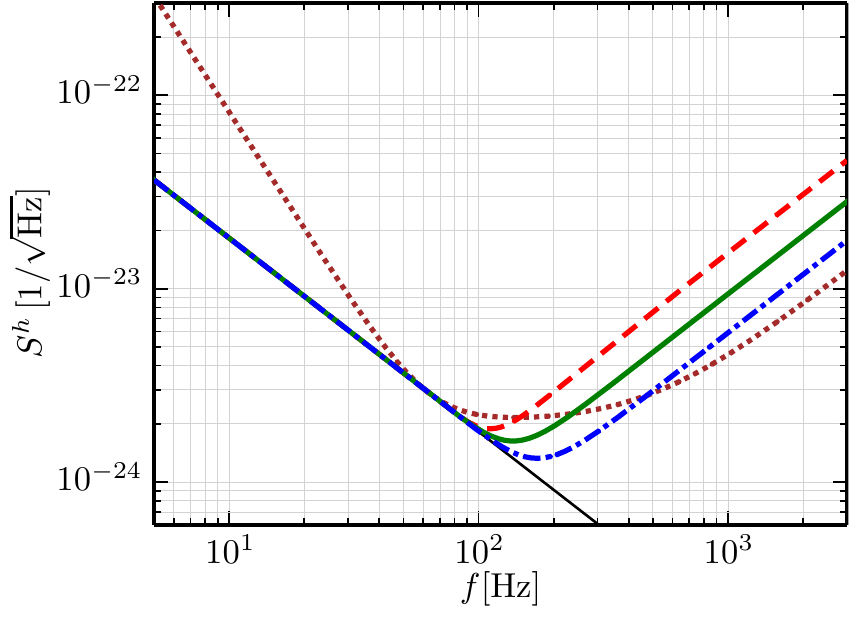}
  \caption{Plots of the total quantum noise spectral density in the double antisymmetric carriers regime without squeezing (dashes), with 6\,db squeezing (solid), and with 12\,db squeezing (dash-dots). The parameters $\varGamma$, $\zeta$, $\beta$, and $\theta$ are given by Eqs.\,\eqref{Gamma_subopt}, \eqref{zeta_subopt}, and Table\,\ref{tab:opt_beta_theta}, respectively. Dots: the baseline interferometer \eqref{S_sum_base}, at $\gamma=2\pi\times500\,{\rm s^{-1}}$ (dots).  Thin solid line: the SQL \eqref{S_SQL}. In all cases, $J=(2\pi\times100)^3\,{\rm s^{-3}}$ and $\eta=1$ (no losses).}\label{fig:1pair}
\end{figure}

Examples of the resulting total quantum noise spectral densities, based on the simplified analytical optimization procedure, described in App.\,\ref{app:suboptimal_lf}, are shown in Fig.\,\ref{fig:1pair}. Comparison of Figs.\,\ref{fig:baseline} and \ref{fig:1pair} shows (assuming the Advanced LIGO parameters), that the double-carrier Michelson/Fabry-P\'erot interferometer can provide the sensitivity comparable with the one of the simplified Sagnac interferometer with uncorrelated quantum noises described by Eq.\,\eqref{S_sum_SM_nocorr}.

We would like to emphasize also the unusual dependence of the quantum noise on the circulating optical power and the squeezing power in the double antisymmetric carriers regime. Similar to the the ordinary single-carrier Michelson/Fabry-P\'erot interferometer case and to the quantum speedmeter one, the high-frequency noise spectral density decreases with the power and the squeezing increase, albeit the dependence is different: $(I_ce^r)^{-4/3}$ [see Eq.\,\eqref{S_lf_scaling}] instead of $(I_ce^{2r})^{-1}$. On contrary to these cases, the low-frequency noise, after the proper adjustment of the parameters $\varGamma$, $\zeta$, $\beta$, and $\theta$, does not increase. Therefore, the double antisymmetric carriers regime does not require the frequency-dependent squeezing or the variational readout to take full advantage of the stronger optical power and/or squeezing.


\subsection{Single interferometer xylophone}\label{ssec:mult_carriers}

The effective shot noise spectral density in the antisymmetric double carrier regime [see Eqs.\,(\ref{S_xx_eff2}, \ref{S_xx_SM})] has one minimum at the frequency $\Omega_0\propto\varGamma$ [see Eq.\,\eqref{Omega_0}], with the width depending on $\beta$ and the squeezing power $e^{2r}$, see App.\,\ref{app:suboptimal_nb}. At lower and at higher frequencies, this spectral density increases as $1/\Omega^2$ and as $\Omega^2$, respectively. The corresponding effective radiation pressure noise spectral density \eqref{S_FF_eff2} mirrors this frequency dependence, having the maximum at $\Omega_0$ and decreasing as $\Omega^2$ and as $1/\Omega^2$ at lower and higher frequencies, respectively.

Therefore, several pairs of the antisymmetric (or nearly antisymmetric) carriers tuned to different values of $\Omega_0$ can be combined together to form a xylophone-like configuration, with each of the pairs responsible for its own frequency band.  Varying parameters of the pairs, it is possible to flexibly shape the resulting total quantum noise spectral density, described by Eqs.\,(\ref{eq:s_sum_1}, \ref{S_eff}).

In particular, the high-frequency sensitivity of the antisymmetric double carrier regime can be improved by adding one or more additional pair(s) of carriers tuned to higher frequencies than the main one. Evidently, in the scenario with the limited total circulating optical power, a part of this power has to be relocated from the first pair to the additional ones, degrading its sensitivity. However, estimates show, that this degradation is more than compensated by the additional pairs and the overall sensitivity improves with the increase of the number of pair.

\begin{figure}
  \includegraphics{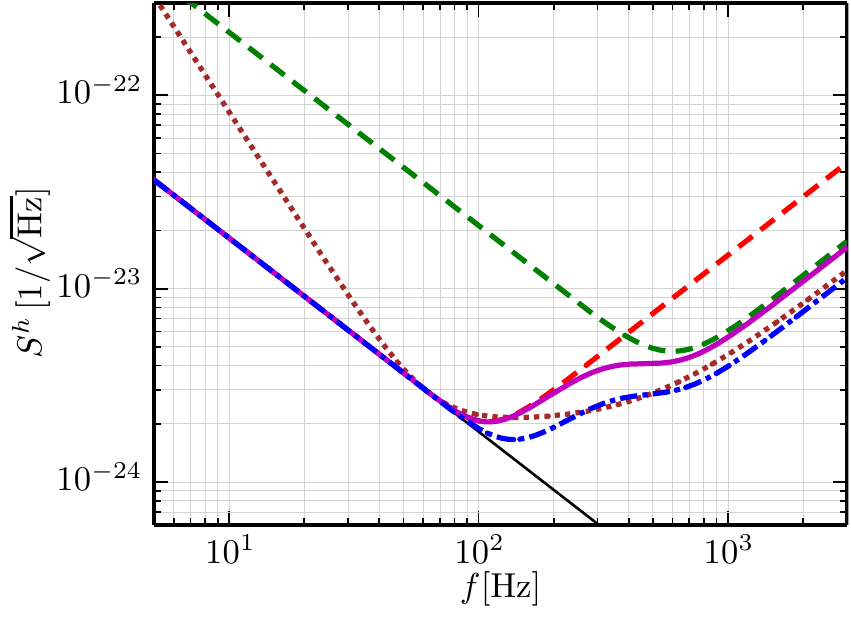}
  \caption{Plots of the total quantum noise spectral densities of the xylophone configuration with two pairs of antisymmetric carriers with 6\,db (solid) and with 12\,db (dash-dots) squeezing. The values of $\varGamma$ are given by Eq.\,\eqref{Gamma_subopt} the lows-frequency pair and Eq.\,\eqref{Gamma_subopt_hf} with $\Omega_0=2\pi\times600\,{\rm Hz}$ for the high-frequency pair. The parameters $\zeta$, $\beta$, and $\theta$ are given by Eq.\,\eqref{zeta_subopt}, and Table\,\ref{tab:opt_beta_theta}, respectively. The optical power is distributed evenly between the all carriers. Dashes: the total quantum noise spectral densities of the individual pairs. Dots: the baseline interferometer \eqref{S_sum_base}, at $\gamma=2\pi\times500\,{\rm s^{-1}}$ (dots). Thin solid line: the SQL \eqref{S_SQL}. In all cases, the total circulating optical power corresponds to $J=(2\pi\times100)^3\,{\rm s^{-3}}$ and $\eta=1$ (no losses).}\label{fig:2pairs}
\end{figure}

An example of the configuration with two pairs of antisymmetric carriers (four carriers total, with the optical power evenly distributed among them) is shown in Fig.\,\ref{fig:2pairs}. Parameters of the low-frequency component are calculated using the same optimization procedure, that was used for the previous example (see App.\,\ref{app:suboptimal_lf}). For the high-frequency pair, another procedure was used, see App.\,\ref{app:suboptimal_bb}, which does not take into consideration the radiation pressure noise, which in this case is negligibly small, but takes into account instead, that the minimum of the shot noise spectral density has to correspond to some given frequency $\Omega_0$.

The total noise spectral density of the higher-frequency pair in this case scales with the optical power and with the squeezing power as $(I_ce^r)^{-1}$ (a bit weaker, than in the previous case).

\begin{figure}
  \includegraphics{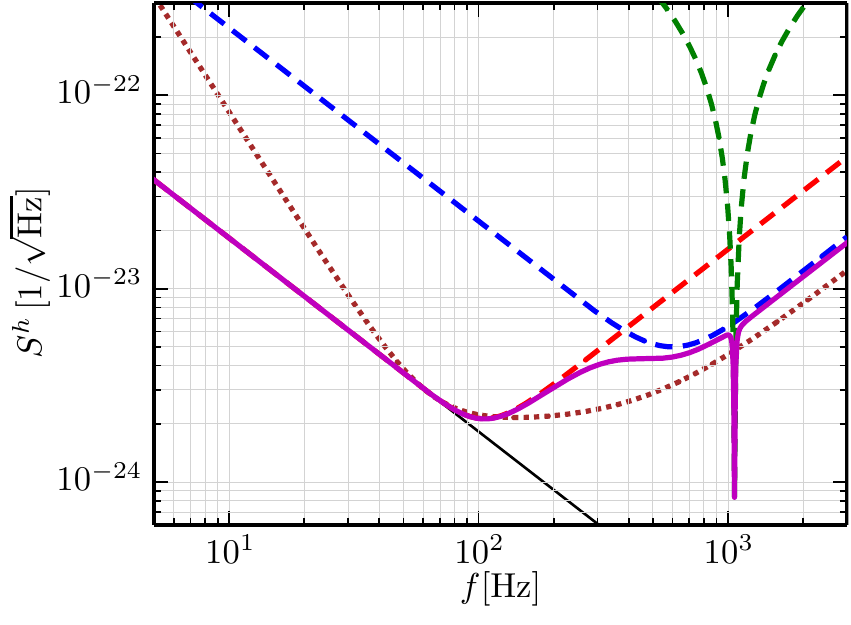}
  \caption{Solud: plot of the total quantum noise spectral densities of the xylophone configuration with two broadband pairs of antisymmetric carriers, with the parameters defined in the same way as in Fig.\,\ref{fig:2pairs}, and one additional narrow-band pair with $\varGamma=4\pi\times532.7\,{\rm s}^{-1}$ (the double frequency of the pulsar J0034-0534), $\beta = \pi/2 - 0.002$, $\theta =  \pi/2$. The optical power is distributed among the all carriers as 45\%:45\%:10\%, and 6\,db squeezing is used for all carriers. Dashes: the total quantum noise spectral densities of the individual pairs. Dots: the baseline interferometer \eqref{S_sum_base}, at $\gamma=2\pi\times500\,{\rm s^{-1}}$ (dots). Thin solid line: the SQL \eqref{S_SQL}. In all cases, the total circulating optical power corresponds to $J=(2\pi\times100)^3\,{\rm s^{-3}}$ and $\eta=1$ (no losses).}\label{fig:2+1pairs}
\end{figure}

The xylophone configuration can also be used to create ``on demand'' some special features of the quantum noise spectral density, for example, narrow-band minima at some given frequencies, associated with the known pulsars. This possibility is demonstrated in Fig.\,\ref{fig:2+1pairs}, where the total quantum noise of a configuration with three antisymmetric pairs is shown. The parameters  of the first two (broadband) pairs are optimized in the same way as in the previous example. However, 10\% of the total optical power is relocated to the third narrow band pair. Parameters of this pair are calculated using the optimization procedure described in App.\,\ref{app:suboptimal_nb}. As an example of millisecond pulsars, we have chosen J0034-0534 \cite{ATNF}, which has the rotation frequency $f_0\approx532.7\,{\rm Hz}$ and therefore presumably radiates near-monochromatic gravitation waves at frequency $2f_0\approx1065.4\,{\rm Hz}$.

\section{Numerical optimization}\label{sec:optimization}

It is evident that the rigorous analytical optimization of the considered above multi-carrier configurations, which takes into account optical losses and various technical noise sources, is impossible. Therefore, here we perform a numerical optimization. As a figure of merit, we use the following cost function \cite{Miao1305_3957}:
\begin{equation}
  \mathcal{C}(\textbf{x}) = \int_{f_{\rm min}}^{f_{\rm max}}
    {\rm log_{10}}\bigl[S_{\rm sum}(2\pi f,\textbf{x}) + S_{\rm tech}(2\pi f)\bigr]
     d({\rm log_{10}}f) \,,
\end{equation}
where $S_{\rm sum}$ is the total  quantum noise spectral density defined by Eq.\,\eqref{S_total}, $S_{\rm tech}$ is the total spectral density of the technical noise  calculated by means of the standard LSC software tool GWINC \cite{GWINCsite}, $f_{\rm min}=5$Hz and $f_{\rm max}=1.5$kHz are the minimal and the maximal frequencies of the optimization procedure, and $\textbf{x}$ is the set of parameters to be optimized. Minimization of this cost function reduces the quantum noise at all frequencies between $f_{\rm min}$ and $f_{\rm max}$ with respect to the technical noise, providing a smooth broadband shape of the total noise spectral density suitable for detection of GW radiation from various types of sources.

The parameters set ${\bf x}$ consists of $2P$ vectors of the form
\begin{equation}
  {\bf x}_j = \{J_j,\delta_j,\gamma_j,\zeta_j,r_j,\theta_j\} \,,
\end{equation}
describing the individual carriers, where $P$ is the number of the carrier pairs. We assume the following relaxed version of the antisymmetry condition \eqref{antisymm}:
\begin{subequations}\label{antisymm_num}
  \begin{align}
    J_{2p-1} &= J_{2p} \,, \\
    r_{2p-1} &= r_{2p} \,, \\
    \varGamma_{2p-1} &= \varGamma_{2p} \,, \\
    \beta_{2p-1} &= -\beta_{2p} \,.
  \end{align}
\end{subequations}
where $p=1\dots P$ is the pair number, varying the homodyne and the squeeze angles $\zeta_j$, $\theta_j$ independently in order to introduce some residual cross-correlation of the shot and the radiation pressure noises. We suppose that the total circulating power is limited by $840\,{\rm kW}$, which is equivalent to $\sum_j J_j \le (2\pi\times100)^3\,{\rm s}^{-3}$, and the squeezing --- by 6\,db ($e^{2r_j}\le4$)


\begin{figure}
  \includegraphics{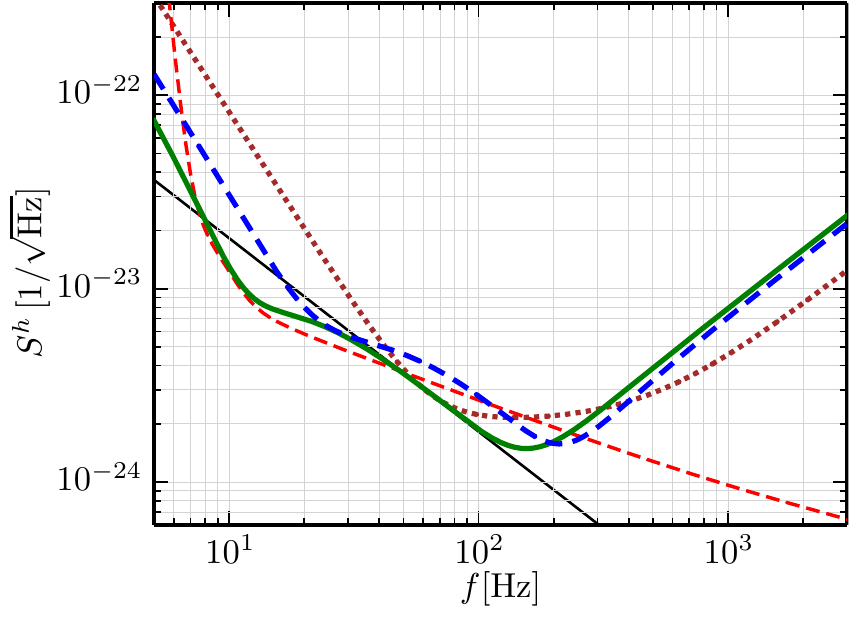} \\
  \includegraphics{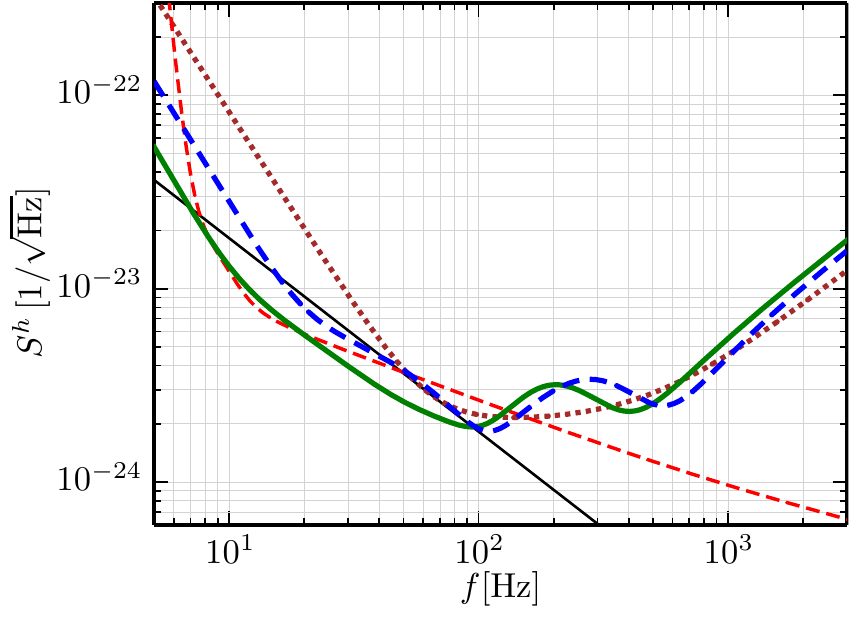}
  \caption{Numerically optimized quantum noise spectral densities for one (top) and two (bottom) pairs of carriers, with $\eta=1$ (solid) and $\eta=0.95$ (dashes). The corresponding optimal parameters are listed in Table \ref{tbl:num_opt}. In all cases,  the total circulating optical power corresponds to $J=(2\pi\times100)^3\,{\rm s^{-3}}$ and 6\,db squeezing is used for all carriers. Dashes: the total quantum noise spectral densities of the individual pairs. Dots: the baseline interferometer \eqref{S_sum_base}, at $\gamma=2\pi\times500\,{\rm s^{-1}}$ (dots). Thin solid line: the SQL \eqref{S_SQL}.
  Thin dashed line: the total technical noise.}\label{fig:num_opt}
\end{figure}


\begin{table*}[t]
  \begin{ruledtabular}
    \begin{tabular}{l|ccccccc|ccccccc}
      & $I_{1,2}$ & $\varGamma_{1,2}$ & $\beta_1=-\beta_2$ &
        $\zeta_1$ & $\zeta_2$ & $\theta_1$ & $\theta_2$ &
        $I_{3,4}$ & $\varGamma_{3,4}$ & $\beta_3=-\beta_4$ &
        $\zeta_3$ & $\zeta_4$ & $\theta_3$ & $\theta_4$ \\
      \hline
      1 pair, $\eta = 1$ & 420\,kW & 550$\,{\rm s}^{-1}$ & -1.0 & -1.12 & 1.14 &
        0.43 & -0.58 & --- & --- & ---& --- & --- & --- & --- \\
      1 pair, $\eta = 0.95$ & 420\,kW & 820$\,{\rm s}^{-1}$ & -1.13 & -1.43 & 1.57 &
        0.16 & -0.15 & --- & --- & ---& --- & --- & --- & --- \\
      \hline
        2 pairs, $\eta = 1$ & 140\,kW & 430$\,{\rm s}^{-1}$ & -1.09 & -1.12 & 1.18 &
        -0.08 & -0.65 &
        280\,kW & 1400$\,{\rm s}^{-1}$ & -0.98 & -1.21 & 1.16 & 0.49 & -0.56 \\
        2 pairs, $\eta = 0.95$ & 145\,kW & 525\,${\rm s}^{-1}$ & -0.915 & -1.41 & 1.56 &
        0.12 & -0.27 &
        265\,kW & 2100$\,{\rm s}^{-1}$ & -0.98 & -1.51 & 1.62 & 0.25 & -0.20 \\
    \end{tabular}
  \end{ruledtabular}
  \caption{The optimized parameters for the one and two pairs of carriers. The optimal total circulating power and the optimal squeezing in all cases are equal to the maximal allowed values of $840\,{\rm kW}$ and 6\,db, respectively.}\label{tbl:num_opt}
\end{table*}

The optimized quantum noise spectral densities are shown in Fig.~\ref{fig:num_opt} and the corresponding optimal parameters are listed in Table~\ref{tbl:num_opt}. Two main conclusions can be drawn from these results.

First, comparison of these spectral densities with the ones of the ideal perfectly antisymmetric regime (see Figs.\,\ref{fig:1pair} and \ref{fig:2pairs}) shows,
that relaxing in some degree the conditions (\ref{zeta_antisymm}, \ref{theta_antisymm}, \ref{zeta_subopt}) and creating thus the cross-correlation of the effective  shot noise and the effective radiations pressure noise, it is possible to push the total quantum noise below the SQL in low frequency band, keeping the high-frequency quantum noise virtually unchanged. The price for this is the quantum noise increase at very low frequencies $f\lesssim10\,{\rm Hz}$. Taking into account, that this frequency band is dominated by the technical noise anyway, this trade-off could improve the overall sensitivity.

Second, it is easy to see, that the multi-carrier regime considered here is sensitive to the optical losses. The reason for this is evident: this regime is heavily relies on the cross-correlations of the shot and radiation pressure noises of the individual optical carriers (see Appendix \ref{app:N_QN}), which are vulnerable to the optical losses.

\section{Discussion}\label{sec:discussion}

Discussing the advantages and disadvantages of the proposed scheme, as well as the prospects of its implementation in GW detectors, we use the frequency-dependent squeezing scheme created by a single relatively short filter cavity \cite{10a1Kh, Evans_PRD_88_022002_2013} as a reference. 

Both schemes promise similar overall sensitivity gain, but ours is more focused on the low-frequency band dominated by the radiation pressure noise and almost no gain at high frequencies. Both share the same main shortcoming, namely, the vulnerability to the optical losses, which is a general feature of methods for overcoming the SQL based on the quantum noise cross-correlation (which includes, in particular, all the filter cavities based schemes, as well as the quantum speedmeter \cite{12a1DaKh}).

Concerning the complexity of practical implementation of the multi-carrier scheme, its most sophisticated element is the output optics which has to spatially separate the output beams and to send each of them to the corresponding homodyne detector. For a single pair, this separation can be implemented by using two orthogonal polarizations for the two carriers, as it was proposed in the initial paper\,\cite{Rehbein2008}. In the case of two and more pairs, the output beams can be separated by means of short (table-top scale) filter cavities. Assuming the following parameters: the length $l_f=1\,{\rm m}$, the losses per bounce $A_f\sim10^{-5}$, and the resulting  quantum inefficiency $1-\eta_f \sim10^{-2}$ \footnote{Defined as $1-\eta_f = \frac{A_f}{T_f+A_f}$, where $T_f$ is the input mirror power transmissivity.}, the half-bandwidth of such a cavity can be estimated as
\begin{equation}
  \gamma_f = \frac{cA_f}{4l_f(1-\eta_f)} \sim 2\pi\times10\,{\rm kHz} \,.
\end{equation}
If detunings between the carriers exceed 100\,kHz, which roughly corresponds to three free spectral ranges of the Advanced LIGO interferometer, then this bandwidth gives the separation efficiency better than 99\%. In order to implement different values of the interferometer bandwidth $\gamma$ for different carriers, the optical outputs can be equipped by the additional signal recycling mirrors, which either supplement the main signal recycling mirror or completely replace it.

Concerning the advantages of the multi-carrier scheme, we would like to name two of them.
First, simple brute-force increase of the circulating optical power and/or the squeezing rate improves high-frequency sensitivity the multi-carrier scheme without degradation of the low-frequency one. In the ``ordinary'' single-carrier Michelson interferometer, increase of the circulating optical power and/or the squeezing improve the high-frequency sensitivity, but degrades the low-frequency one. The filter cavities allow to avoid this degradation, but in this case increase of the circulating power have to be supplemented by the proportional increase of the squeezing in order to keep the low-frequency sensitivity unchanged.

Second, the multi-carrier scheme allows to tune very flexibly the shape of the quantum noise. In particular, using additional carriers pairs, it is possible to create deep minima in the quantum spectral density without affecting the sensitivity at other frequencies.

\acknowledgments

The work of M.K. and F.K. was supported by Russian Foundation for Basic Research grant No.11-02-00383-a. The work of F.K. was supported by LIGO NSF grant PHY-1305863. H.M. has been supported by the Marie Curie research fellowship. The numerical optimization was performed using the computer equipment donated to S. Danilishin by the Alexander von Humboldt Foundation (Germany) within the frames of the Alumni Support Program.

The paper has been assigned LIGO document number P1400163.

\newpage

\appendix

\section{Multi-channel position meter}\label{app:N_QN}

In order to simplify the equations, we use the two-sided force normalized spectral density $S^F_{\rm sum}$ of the sum quantum noise in this Appendix (see details in \cite{12a1DaKh}); the single-sided GW strain signal normalized spectral density used in the main text can be obtained from by means of the following equation:
\begin{equation}
  S_{\rm sum}(\Omega) = \frac{8S^F_{\rm sum}(\Omega)}{L^2M^2\Omega^4} \,.
\end{equation}

Consider a system consisting of $N$ linear meters measuring position $x$ of a test object. Each of the meters is described by its measurement noise $\hat{x}_j$ and back action noise $\hat{F}_j$ ($i=1..N$), with the spectral densities $S_{xx}^{(j)}$, $S_{FF}^{(j)}$, $S_{xF}^{(j)}$. The test object is described by its susceptibility function
\begin{equation}
  \chi(\Omega) = \frac{1}{D(\Omega)} \,,
\end{equation}
with  the possible dynamic back action of the meters (the optical springs) included into it.

In Fourier representation, outputs of these meters are equal to
\begin{equation}
  \mathcal{G}_j(\Omega)
  = G(\Omega) + D(\Omega)\hat{x}_j(\Omega) + \sum_{j=1}^N\hat{F}_j(\Omega) \,,
\end{equation}
where $G$ is the signal force. The combined output is equal to
\begin{equation}
  \mathcal{G}(\Omega) = \sum_{j=1}^N\alpha_j(\Omega)\mathcal{G}_j(\Omega)
  = G(\Omega) + \hat{F}_{\rm sum}(\Omega) \,,
\end{equation}
where $\alpha_j(\Omega)$ are weight functions satisfying the normalization condition
\begin{equation}\label{norm}
  \sum_{j=1}^N\alpha_j(\Omega) = 1
\end{equation}
and
\begin{equation}
  \hat{F}_{\rm sum}(\Omega) = \sum_{j=1}^N
    \bigl[D(\Omega)\alpha_j(\Omega)\hat{x}_j(\Omega) + \hat{F}_j(\Omega)\bigr]
\end{equation}
is the total effective noise force with the spectral density being equal to
\begin{multline}\label{S_sum_s}
  S_{\rm sum}^F(\Omega)  = \sum_{j=1}^N\Bigl\{
      |D(\Omega)|^2|\alpha_j(\Omega)|^2S_{xx}^{(j)}(\Omega) \\
      + 2\Re\bigl[D(\Omega)\alpha_j(\Omega)S_{xF}^{(j)}(\Omega)\bigr]
      + S_{FF}^{(j)}(\Omega)
    \Bigr\} .
\end{multline}

Using the vector notation, Eqs.\,(\ref{norm}, \ref{S_sum_s}) can be rewritten as follows:
\begin{equation}
  {\bf A}^\dagger(\Omega){\bf 1} = 1 \,, \label{norm_v}
\end{equation}
\begin{multline}\label{S_sum_v}
  S^F_{\rm sum}(\Omega)
  = |D(\Omega)|^2{\bf A}^\dagger\mathbb{S}_{xx}(\Omega){\bf A}(\Omega) \\
    + 2\Re\bigl[D(\Omega){\bf A}^\dagger(\Omega){\bf S}_{xF}(\Omega)\bigr]
    + \sum_{j=1}^NS_{FF}^{(j)}(\Omega) \,,
\end{multline}
where
\begin{gather}
  {\bf A}^\dagger(\Omega)
    = \begin{pmatrix} \alpha_1(\Omega) & \dots & \alpha_N(\Omega) \end{pmatrix} , \\
  {\bf 1} = \begin{pmatrix} 1 \\ \vdots \\ 1 \end{pmatrix} , \\
  {\bf S}_{xF}(\Omega)
    = \begin{pmatrix} S_{xF}^{(1)}(\Omega) \\ \vdots \\ S_{xF}^{(N)}(\Omega) \end{pmatrix}
      , \\
  \mathbb{S}_{xx} = \begin{pmatrix}
      S_{xx}^{(1)}(\Omega) & & \text{\raisebox{-0.8ex}{\Large 0}} \\[-1ex]
      & \ddots & \\[-1ex]
      \text{\Large 0} & & S_{xx}^{(N)}(\Omega)
    \end{pmatrix} .
\end{gather}
With account of condition \eqref{norm_v}, minimum of \eqref{S_sum_v} is given by
\begin{equation}
  {\bf A}^\dagger(\Omega)
  = -\frac{\lambda{\bf 1}^\dagger + D^*(\Omega){\bf S}_{xF}^\dagger(\Omega)}
      {|D(\Omega)|^2}\,
      \mathbb{S}_{xx}^{-1}(\Omega)\,.
\end{equation}
where $\lambda$ is the Lagrange factor defined by \eqref{norm_v}:
\begin{equation}
  \lambda = -\frac{
      |D(\Omega)|^2
      + D^*(\Omega){\bf S}_{xF}^\dagger(\Omega)\mathbb{S}_{xx}^{-1}(\Omega){\bf 1}
    }{{\bf 1}^\dagger\mathbb{S}_{xx}^{-1}(\Omega){\bf 1}}
\end{equation}
Therefore (returning back to the scalar notation),
\begin{equation}\label{alpha_j_opt}
  \alpha_j(\Omega) = \frac{1}{S_{xx}^{(j)}(\Omega)}\left\{
      S_{xx}^{\rm eff}(\Omega)
      + \frac{\bigl[S_{xF}^{\rm eff}(\Omega) - S_{xF}^{(j)}(\Omega)\bigr]^*}{D(\Omega)}
    \right\}
\end{equation}
and
\begin{multline}\label{S_total}
  S^F_{\rm sum}(\Omega) = |D(\Omega)|^2S_{xx}^{\rm eff}(\Omega)
    + 2\Re\bigl[D(\Omega)S_{xF}^{\rm eff}(\Omega)\bigr] \\
    + S_{FF}^{\rm eff}(\Omega)  \,,
\end{multline}
where
\begin{subequations}\label{S_eff}
  \begin{gather}
    S_{xx}^{\rm eff}(\Omega)
      = \left[\sum_{j=1}^N\frac{1}{S_{xx}^{(j)}(\Omega)}\right]^{-1} , \label{S_eff_S_x}\\
    S_{FF}^{\rm eff}(\Omega)
      = \sum_{j=1}^N\left[
            S_{FF}^{(j)}(\Omega)
            - \frac{|S_{xF}^{(j)}(\Omega)|^2}{S_{xx}^{(j)}(\Omega)}
          \right]
        + \frac{|S_{xF}^{\rm eff}(\Omega)|^2}{S_{xx}^{\rm eff}(\Omega)} \,, \\
    S_{xF}^{\rm eff}(\Omega)
      = S_{xx}^{\rm eff}(\Omega)
          \sum_{j=1}^N\frac{S_{xF}^{(j)}(\Omega)}{S_{xx}^{(j)}(\Omega)}
  \end{gather}
\end{subequations}
are the effective quantum noise spectral densities.

It follows from these equations, that
\begin{multline}
  S_{xx}^{\rm eff}(\Omega)S_{FF}^{\rm eff}(\Omega) - |S_{xF}^{\rm eff}(\Omega)|^2 \\
  = S_{xx}^{\rm eff}(\Omega)\sum_{j=1}^N\left[
      S_{FF}^{(j)}(\Omega)
      - \frac{|S_{xF}^{(j)}(\Omega)|^2}{S_{xx}^{(j)}(\Omega)}
      \right]
  \ge \frac{\hbar^2}{4} \,.
\end{multline}
Therefore, if for all $j$ the exact equality takes place in the uncertainty relation \eqref{eq:uncert}, then the same is valid for the effective spectral densities:
\begin{equation}\label{uncert_eff}
  S_{xx}^{\rm eff}(\Omega)S_{FF}^{\rm eff}(\Omega) - |S_{xF}^{\rm eff}(\Omega)|^2
  = \frac{\hbar^2}{4} \,.
\end{equation}

\section{Quantum noise of the laser interferometric position meter}\label{app:SxSFSxF}

In this Appendix, we consider the single carrier features and therefore omit for brevity the indices enumerated the carriers.

\subsection{General equations}

Neglecting for simplicity the intra-cavity optical losses in comparison with the optical losses in the output optical elements and the photodetectors quantum inefficiency (which can be considered as frequency-independent ones), the quantum noise spectral densities and the optical rigidity of the laser interferometric position meter can be presented as follows (derivation of these equations can be found in \cite{12a1DaKh}):
\begin{subequations}\label{SxSFSxF}
  \begin{multline}\label{S_x_PM}
    S_{xx} = \frac{\hbar}{4MJ\gamma}\,
      \frac{1}{\varGamma^2\sin^2(\zeta-\beta) + \Omega^2\sin^2\zeta} \\ \times
      \biggl[
          Q_c^2(\Omega)e^{2r} + Q_s^2(\Omega)e^{-2r}
          + \frac{1-\eta}{\eta}|\mathcal{D}(\Omega)|^2
        \biggr] ,
  \end{multline}
  \begin{equation}\label{S_F_PM}
     S_{FF} = \frac{\hbar MJ\gamma}{|\mathcal{D}(\Omega)|^2}
      \Bigl[|P_c(\Omega)|^2e^{2r} + |P_s(\Omega)|^2e^{-2r}\Bigr] ,
  \end{equation}
  \begin{multline}\label{S_xF_PM}
    S_{xF} = \frac{\hbar}{2\mathcal{D}^*(\Omega)}\,
      \frac{Q_c(\Omega)P_c(\Omega)e^{2r} + Q_s(\Omega)P_s(\Omega)e^{-2r}}
        {\varGamma\sin(\zeta-\beta) - i\Omega\sin\zeta} \,,
  \end{multline}
\end{subequations}
\begin{equation}\label{K}
  K(\Omega) = \frac{MJ\delta}{\mathcal{D}(\Omega)} \,,
\end{equation}
where
\begin{subequations}
  \begin{align}
    Q_c(\Omega) &= \varGamma^2\cos(2\beta+\theta-\zeta) + \Omega^2\cos(\theta-\zeta)\,,\\
    Q_s(\Omega) &= -\varGamma^2\sin(2\beta+\theta-\zeta) - \Omega^2\sin(\theta-\zeta)\,,\\
    P_c(\Omega) &= \varGamma\cos(\theta+\beta) + i\Omega\cos\theta \,, \\
    P_s(\Omega) &= -\varGamma\sin(\theta+\beta) - i\Omega\sin\theta \,,
  \end{align}
\end{subequations}
and
\begin{equation}
  \mathcal{D}(\Omega) = (\gamma-i\Omega)^2 + \delta^2 \,.
\end{equation}

\subsection{Speedmeter-like frequency dependence of the shot noise}

Consider the ultimate case of the condition \eqref{low_freq}, assuming that
\begin{equation}\label{zeta_subopt}
  \zeta=\beta \,.
\end{equation}
This assumption gives the exact speedmeter-like frequency dependence of the shot noise:
\begin{multline}\label{S_xx_SM}
  S_{xx}(\Omega) = \frac{\hbar}{4MJ\varGamma\cos\beta\sin^2\beta} \\ \times
    \frac{A\varGamma^4 + 2B\varGamma^2\Omega^2 + C\Omega^4}{\Omega^2} \,,
\end{multline}
where
\begin{subequations}
  \begin{align}
    A &= e^{2r}\cos^2(\beta+\theta) +e^{-2r}\sin^2(\beta+\theta) \,, \\
    B &= e^{2r}\cos(\beta+\theta)\cos(\theta-\beta) \nonumber \\
      &\hspace*{4em} + e^{-2r}\sin(\beta+\theta)\sin(\theta-\beta) \,, \\
    C &= e^{2r}\cos^2(\theta-\beta) +e^{-2r}\sin^2(\theta-\beta) \,.
  \end{align}
\end{subequations}
The low- and high-frequency asymptotics of \eqref{S_xx_SM} are equal to
\begin{subequations}
  \begin{align}
    S_{xx}(\Omega\to0) &= \frac{\hbar\varGamma^3}{4MJ\Omega^2\cos\beta\sin^2\beta}\,A \,,
      \label{S_xx_lf} \\
    S_{xx}(\Omega\to\infty)
      &=\frac{\hbar\Omega^2}{4MJ\varGamma\cos\beta\sin^2\beta}\,C\,,\label{S_xx_hf}
  \end{align}
\end{subequations}
The minimum of \eqref{S_xx_SM} is equal to
\begin{equation}
  S_{xx}(\Omega_0) = \frac{\hbar\varGamma}{2MJ\cos\beta\sin^2\beta}\,(\sqrt{AC} + B) \,,  \label{S_xx_mf}
\end{equation}
where
\begin{equation}\label{Omega_0}
  \Omega_0 = \varGamma\left(\frac{A}{C}\right)^{1/4} .
\end{equation}

\section{Sub-optimal regimes of the dual carrier interferometer}\label{app:suboptimal}

Here we analytically calculate a sub-optimal parameters values of the antisymmetric dual-carrier regime which we use in the plots in Sec.\,\ref{sec:idea}. We enumerate the carriers by the index $j$, assuming the condition \eqref{antisymm} for the odd and the even components.

\subsection{One pair of carriers or low-frequency pair of the xylophone}\label{app:suboptimal_lf}

Start with requirement, that the low frequency asymptotic of the total quantum noise spectral density has to be equal to the SQL:
\begin{equation}
  S_{\rm sum}(\Omega\to0) = \frac{\hbar}{M\Omega^2} \,.
\end{equation}
With account of Eqs.\,(\ref{eq:s_sum_1}, \ref{antisymm}, \ref{S_2_carriers}, \ref{S_xx_lf}), it gives:
\begin{multline}\label{Gamma_subopt}
  S_{xx}^{\rm eff}(\Omega\to0) = \frac{S_{xx}^{(j)}(\Omega\to0)}{2}
    = \frac{\hbar}{2M\Omega^2} \hence \\
  \varGamma_j = \left(\frac{4J_j\cos\beta_j\sin_j^2\beta}{A_j}\right)^{1/3} \,,
\end{multline}
where $j=1,2$.

The corresponding high-frequency asymptotic of the total quantum noise is equal to
\begin{multline}
  S_{\rm sum}(\Omega\to\infty) \approx \frac{S_{xx}^{(j)}(\Omega\to\infty)}{2} \\
  = \frac{\hbar\Omega^2}{2M(4J_j)^{4/3}}\,F^{1/3}(\beta_j,\theta_j) \,,
\end{multline}
where
\begin{equation}
  F(\beta,\theta) = \frac{AC^3}{\cos^4\beta\sin^8\beta} \,.
\end{equation}
The values of $\beta$ and $\theta$ which provide the minimum of this function are shown in Table\,\ref{tab:opt_beta_theta} for some characteristic values of squeezing. Note that in all cases, $\theta\approx\pi/2+\beta$, which cancels the term proportional to $e^{2r}$ in $A$, giving
\begin{equation}
  F(\beta,\theta) \propto e^{-4r} \,.
\end{equation}
Therefore,  the high-frequency part of the total noise scales with the power and with the squeezing as follows:
\begin{equation}\label{S_lf_scaling}
  S_{\rm sum}(\Omega\to\infty) \propto \frac{1}{(Je^r)^{4/3}} \,.
\end{equation}

\begin{table}
  \begin{ruledtabular}
  \begin{tabular}{cccc}
    $e^{2r}$ & $\varGamma/\Omega_0$ & $\beta$ & $\theta$ \\
    \hline
      1.0  (0\,db) &  1.0  & $-\arccos(1/\sqrt{3})$ &  ---   \\
      2.0  (3\,db) &  0.75  & -1.02 &  0.51 \\
      4.0  (6\,db) &  0.54  & -1.04 &  0.52 \\
     10.0 (10\,db) &  0.34  & -1.05 &  0.52 \\
     $>10.0$       & $e^{-r}/\sin2\beta$ & -1.047 & $\pi/2+\beta$
  \end{tabular}\caption{Values of $\varGamma$, $\beta$, and $\theta$ which minimize function \eqref{ac}}\label{tab:opt_beta_theta}
  \end{ruledtabular}
\end{table}

\subsection{Higher-frequency components of xylophon}\label{app:suboptimal_bb}

At high frequency, the radiation pressure noise can be neglected. In this case, our goal is to get the most broadband shot noise spectral density centered at some given frequency $\Omega_0$. Therefore, we minimize the product of the low- and high-frequency asymptotics
\begin{multline}\label{ac}
  S_{xx}(\Omega\to0)^{\rm eff}\times S_{xx}^{\rm eff}(\Omega\to\infty) \\
  = \frac{1}{4}S_{xx}^{(j)}(\Omega\to0)\times S_{xx}^{(j)}(\Omega\to\infty) \\
  = \frac{1}{4}\left(\frac{\hbar}{4MJ_j}\right)^2
      \frac{\varGamma_j^2A_jC_j}{\cos^2\beta_j\sin^4\beta_j}
\end{multline}
where $j=\{2p+1, 2p+2\}$ and $p=2,\dots$ is the pair number, in $\varGamma_j$, $\beta_j$, and $\theta_j$ for a given value of $\Omega_0$:

Eq.\,\eqref{Omega_0} gives $\varGamma_j$:
\begin{equation}\label{Gamma_subopt_hf}
  \varGamma_j = \Omega_0\left(\frac{C_j}{A_j}\right)^{1/4} \,.
\end{equation}
Therefore,
\begin{equation}
  S_{xx}^{(j)}(\Omega\to0)\times S_{xx}^{(j)}(\Omega\to\infty)
  = \left(\frac{\hbar}{4MJ_j}\right)^2\sqrt{F(\beta_j,\theta_j)} \,,
\end{equation}
with the same optimal values of $\beta_j$ and $\theta_j$ as for the low-frequency pair.

In this case, the noise scales with the power and with the squeezing as follows:
\begin{equation}\label{S_bb_scaling}
  S_{\rm sum}(\Omega\to\infty) \propto \frac{1}{Je^r} \,.
\end{equation}

\subsection{Narrowband optimization}\label{app:suboptimal_nb}

The minimum of \eqref{S_xx_mf} in $\theta$ is provided by
\begin{equation}
  \theta_j = \frac{\pi}{2} \,.
\end{equation}
In this case,
\begin{multline}
  S_{xx}^{(j)}(\Omega) = \frac{\hbar}{4MJ_j\varGamma_j\Omega^2\cos\beta_j\sin^2\beta_j}
    \,\\ \times
    \Bigl[
        (\Omega^2-\varGamma_j^2)^2e^{2r}\sin^2\beta_j
        + (\Omega^2+\varGamma_j^2)^2e^{-2r}\cos^2\beta_j
      \Bigr] .
\end{multline}
If
\begin{equation}
  \left|\alpha_j = \frac{\pi}{2} - \beta_j\right| \ll 1 \,,
\end{equation}
then this spectral density has a sharp minimum at $\Omega=\varGamma_j$. In this case
\begin{equation}
  S_{xx}^{(j)}(\Omega_0+\nu) \approx \frac{\hbar}{MJ\varGamma_j\alpha_j}\,
    \Bigl(\nu^2e^{2r_j} + \varGamma^2\alpha_j^2e^{-2r_j}\Bigr) .
\end{equation}
Therefore, the value of the minimum and its width are equal to
\begin{gather}
  S_{xx}(\varGamma) \approx \frac{\hbar\varGamma_j\alpha_j e^{-2r_j}}{MJ}\,, \\
  \Delta\Omega = 2\varGamma_j\alpha_je^{-2r_j} \,.
\end{gather}


\end{document}